\documentclass[aps,floats,floatfix,twocolumn]{revtex4}

\usepackage{amsfonts,amsmath,amssymb,hyperref}
\usepackage[pdftex]{graphicx}
\usepackage{color}
\usepackage{times}

\begin{document}

\title{Deciphering the global organization of clustering in real complex networks}

\author{Pol Colomer-de-Sim\'on$^{1}$, M. \'Angeles Serrano$^{1}$, Mariano G. Beir\'o$^{2}$, J. Ignacio Alvarez-Hamelin$^{2}$, \& Mari{\'a}n Bogu{\~n}{\'a}$^{1}$}

\affiliation{$^1$Departament de F{\'\i}sica Fonamental, Universitat de Barcelona, Mart\'{\i} i Franqu\`es 1, 08028 Barcelona, Spain}
\affiliation{$^2$INTECIN (CONICET--U.B.A.), Facultad de Ingenier\'ia, Universidad de Buenos Aires, Paseo Col\'on
850, C1063ACV Buenos Aires, Argentina}

\date{\today}

\begin{abstract}
We uncover the global organization of clustering in real complex networks. As it happens with other fundamental properties of networks such as the degree distribution, we find that 
real networks are neither completely random nor ordered with respect to clustering, although they tend to be closer to maximally random architectures. We reach this conclusion by comparing the global structure of clustering in real networks with that in maximally random and in maximally ordered clustered graphs. The former are produced with an exponential random graph model that maintains correlations among adjacent edges at the minimum needed to conform with the expected clustering spectrum; 
the later with a random model that arranges triangles in cliques inducing highly ordered structures. 
To compare the global organization of clustering in real and model networks, we compute $m$-core landscapes, where the $m$-core is defined, akin to the $k$-core, as the maximal subgraph with edges participating at least in $m$ triangles. This property defines a set of nested subgraphs that, contrarily to $k$-cores, is able to distinguish between hierarchical and modular architectures. 
To visualize the $m$-core decomposition we developed the LaNet-vi 3.0 tool.
\end{abstract}

\maketitle

\section{Introduction}

The architecture of real complex systems lay at the midpoint between order and disorder, although its precise location is quite difficult to determine. Disorder in complex networks is manifested by the small-world effect~\cite{Watts:1998ga} and a highly heterogeneous degree distribution~\cite{Barabasi:1999ha}, both properties commonly present in real complex networks~\cite{Dorogovtsev:2003ti,newmanbook}. Order is, on the other hand, manifested by the presence of triangles --or clustering-- representing three point correlations in the system. Indeed, the very concept of order is typically related to the existence of a metric structure in the system which, from the network perspective, is captured by clustering, the smallest network motif able to encode the triangle inequality. Yet, unlike the small-world effect and the heterogeneity of nodes' degrees, clustering is not an emergent property spontaneously generated by paradigmatic connectivity principles such as preferential attachment and, therefore, calls for specific mechanisms for explaining its emergence, thus giving important insights into the nature of network formation and network evolution.

On the other hand, the effects of clustering on the structural and dynamical properties of networks have not yet been conclusively elucidated. In fact, several studies have reported apparently contradictory results concerning the effects of clustering on the percolation properties of networks and little is known on its effects on dynamical processes running on networks~\cite{Serrano:2006dq,Serrano:2006ka,Trapman2007160,Newman:2009fk,Gleeson:2009vn,Gleeson:2009ys,Gleeson:2010zr}. This is further hindered by the technical difficulties of any analytical treatment. Indeed, the presence of strong clustering invalidates, in general, the ``locally tree-like'' assumption used in random graphs, leaving little room for any theoretical study. In an effort to overcome these problems, a new class of clustered network models has been proposed. These models start by defining a certain set of cliques (fully connected subgraphs) of different sizes that are afterwords connected in a random fashion. In this way, by considering cliques as super-nodes, the network connecting these super-nodes is locally tree-like, thus allowing for an analytical treatment~\cite{Trapman2007160,Newman:2009fk,Gleeson:2009vn,Gleeson:2009ys,Karrer:2010nx,Gleeson:2010zr,Allard:2012}. Then, it is possible to generate networks with a given degree distribution $P(k)$ and degree-dependent clustering coefficient $\bar{c}(k)$, defined as the average fraction of triangles attached to nodes of degree $k$.

While this is indeed a fair approach to the problem, triangles generated by these models are arranged in a very specific way, with strong correlations between the properties of adjacent edges. In some sense, we can consider this class of models as generators of maximally ordered clustered graphs. At the other side of the spectrum, we can define an ensemble of maximally random clustered graphs such that correlations among adjacent edges are the minimum needed to conform with the degree-dependent clustering coefficient, but no more. These two types of models define --in a non-rigorous way-- two extremes of the phase space of possible graphs with given $P(k)$ and $\bar{c}(k)$. A simple question arises then: where are real networks positioned in this phase space? To give an answer to this question, we need to go beyond the local properties of networks and to study their global organization. In this paper, we study the global structure of clustering in real networks and compare them with the global structure of clustering induced by the two types of models with identical local properties. More specifically, we analyze the organization of real and model networks into $m$-cores, defined as maximal subgraphs with edges participating at least in $m$ triangles, that is able to distinguish between hierarchical and modular architectures. Interestingly enough, real networks tend to be closer to maximally random clustered graphs, although clear differences are evident. 

\section{Results}

In this paper, we analyze three real paradigmatic networks from different domains: the Internet at the Autonomous System level~\cite{Boguna:2010uq}, the web of trust of the Pretty Good Privacy protocol (PGP)~\cite{Boguna:2004jx}, and the metabolic network of the bacterium {\it E. coli}~\cite{Serrano:2012we}. However, the results obtained here also hold for a wide spectrum of systems (See Supplementary Information for the analysis of a larger set of systems). We first describe their random counterparts, namely, maximally ordered and maximally random clustered graphs with the same degree distribution and clustering spectrum.

\subsection{Network models}
\begin{table*}
\caption{Statistics of real networks and their random counterparts. $N$ is the number of nodes, $E$ is the number of edges, $C$ is the average clustering coefficient averaged only over nodes with degrees $k \ge 2$. We also show the number of disconnected components and the relative size of the giant component.}
\begin{center}
\begin{tabular}{l|c|c|c|c|c|}
  & $N$ & $E$ & $C$ & \# of clusters & Giant component\\ \hline \hline
Internet & 23752 & 58416 & 0.61 & 3 & 99.98\%\\ \hline
Internet clique-based model& 23800$\pm$200 & 50000$\pm$10000& 0.62$\pm$0.01& 2200$\pm$400 & (75$\pm$4)\%\\ \hline 
Internet random $\bar{c}(k)$& 23752 & 58416 & 0.61 & 16$\pm$4 & (99.84$\pm$0.06)\%\\ \hline
Internet random $\bar{c}(k)$, $P(k,k')$ & 23752 & 58416 & 0.61 & 4$\pm$1 & (99.96$\pm$0.02)\%\\ \hline \hline
PGP & 57243 & 61837 & 0.50 & 16221 & 18.65\%\\ \hline
PGP clique-based model& 62000$\pm$1000 & 57200$\pm$200 & 0.506$\pm$0.005 & 13700$\pm$200 & (37$\pm$1) \%\\ \hline
PGP random $\bar{c}(k)$& 57243 & 61837 & 0.487$\pm$0.001 & 15550$\pm$60 & (21.3$\pm$0.4)\%\\ \hline
PGP random $\bar{c}(k)$, $P(k,k')$ & 57243 & 61837 & 0.493$\pm$0.001 & 15810$\pm$20 & (22.3$\pm$0.3)\%\\ \hline \hline
E. Coli & 1010 & 3286 & 0.48 & 2 & 99.8\%\\ \hline
E. Coli clique-based model& 1010$\pm$40 & 3300$\pm$700 & 0.51$\pm$0.01 & 7$\pm$3 & (97.9$\pm$0.6) \%\\ \hline
E. Coli random $\bar{c}(k)$& 1010 & 3286 & 0.48 & 2.2$\pm$0.9 & (99.7$\pm$0.3)\%\\ \hline
E. Coli random $\bar{c}(k)$, $P(k,k')$ & 1010 & 3286 & 0.48 & 7$\pm$2 & (98.2$\pm$0.6)\%\\ \hline \hline
\end{tabular}
\end{center}
\label{table:1}
\end{table*}

One of the best clique-based models to generate maximally ordered clustered networks is the one introduced by Gleeson in~\cite{Gleeson:2009vn}. In this model, nodes belong to single cliques and are also given a number of connections outside their cliques. Then, cliques are considered as super-nodes, each with an effective degree given by the sum of all the external links of the members of the clique, and connected using the standard configuration model. The input of the model is the joint distribution $\gamma(c,k)$, defined as the probability that a randomly chosen node has degree $k$ and belongs to a clique of size $c$. Both the degree distribution and the degree-dependent clustering coefficient are related to function $\gamma(c,k)$. Therefore, by properly choosing its form, it is possible to match the desired degree distribution and clustering. Note, however, that since we start with cliques and not nodes, the number of nodes and their actual degrees are not fixed {\it a priori}. As a consequence, in finite heterogeneous networks, there may be some unavoidable discrepancies between real and random versions of the network. Hereinafter, we denote this model as ``clique-based model'' (CB).

On the other hand, we generate maximally random clustered networks as an ensemble of exponential 
graphs~\cite{Park:2004vo} with Hamiltonian
\begin{equation}
H=\sum_{k=k_{min}}^{k_c} |\bar{c}^*(k)-\bar{c}(k)|,
\end{equation}
where $\bar{c}(k)$ is the target degree-dependent clustering coefficient and $\bar{c}^*(k)$ is the one corresponding to the current state of the network. This Hamiltonian is minimized by means of simulated annealing coupled to a Metropolis rewiring scheme until the current clustering is close enough to the target one (see Methods Section for further details). Here we use two different rewiring schemes. In the first one~\cite{Maslov:2002wp}, degrees of nodes are preserved after each single rewiring event but correlations between the degrees of connected nodes are either destroyed or brought down to the level of the structural ones~\cite{Boguna:2004eh,Serrano:2007nl}. In the second scheme~\cite{Melnik:2011uq}, rewiring events preserve both the degree distribution and the joint degree-degree distribution of connected nodes, $P(k,k')$, so that degree-degree correlations are fully preserved. Hereinafter, we denote these models as ``maximally random models'' (MR). We would like to stress that, even though there are many models of exponential random graphs generating clustered graphs~\cite{Frank:1986vm,Milo:2002vs,Foster:2010uq}, none of them reproduces the actual clustering spectrum as a function of node degree. In this sense, our maximally random model gets closer to real networks.

Notice that none of the random models used in this paper enforces global connectivity of the network in a single connected component. Therefore, the number of disconnected components and the size of the giant (or largest) component must be considered as predictions of the models, which can be readily compared to those of real networks. In Table~\ref{table:1}, we show this comparison with the networks analyzed in this paper. Quite remarkably, in the case of the Internet, MR models predict the existence of, basically, a single connected component, as it is also observed in the real network. On the other hand, the CB model generates a very large number of disconnected components and a giant component significantly smaller than the real one. Even more surprising are the results for the PGP web of trust. The real network is fragmented into a large number of small components whereas its giant component occupies around 18\% of the network. All models generate a similar number of disconnected components. However, the relative size of the giant component is very well reproduced by MR models, whereas the CB model predicts a giant component twice as large. In the case of the metabolic network of the bacterium {\it E. coli}, all models predicts the existence of a single connected component, in good agreement with the real network.

\subsection{Revealing network hierarchies: $k$-cores and $m$-cores}

\begin{figure}[t]
\centerline{\includegraphics[width=\linewidth]{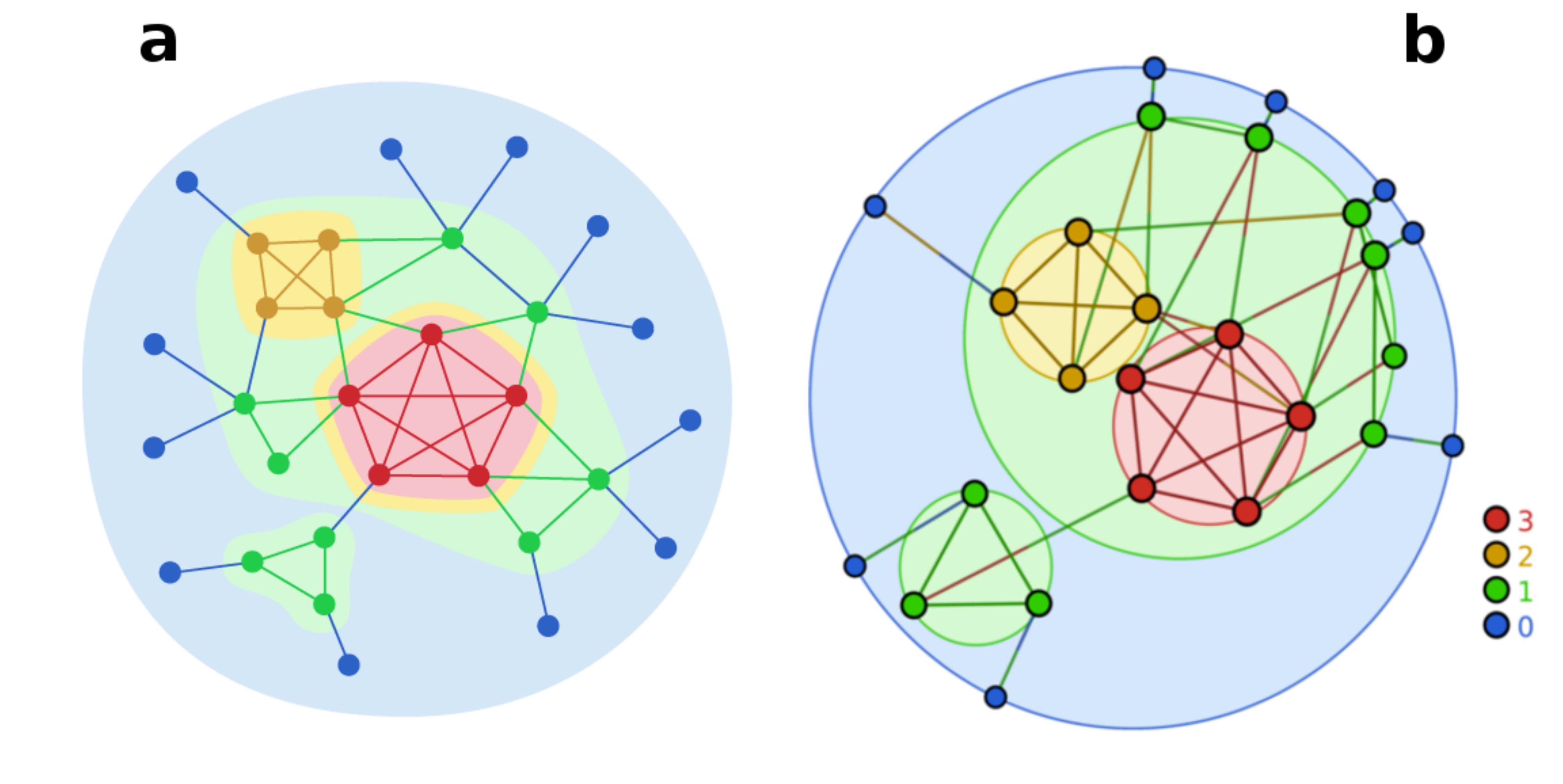}}
\caption{{\bf m-cores decomposition and its visualization.} The example network in {\bf a} is colored according to the $m$-coreness of nodes and edges. Nodes and edges colored in blue belong to the $m0$-core but not to the $m1$-core. Nodes and edges colored in green belong to the $m1$-core but not to the $m2$-core, etc. The same structure is represented in {\bf b} with our visualization tool. The outermost circle in blue represents the $m0$-core, with nodes of $m$-coreness 0 located in its perimeter. The $m1$-core --which is contained within the $m0$-core-- is fragmented in two disconnected components, which are represented as two non-overlapping circles within the outermost one and with nodes of $m$-coreness 1 located in their perimeters. The larger of these two components is further fragmented in two disconnected components representing the $m2$-core and $m3$-core. The angular positions of nodes in each circumference are chosen to minimize the angular separation with their neighbors in different layers. Notice that in this representation, each edge is colored with two colors, corresponding to the colors of the $m$-coreness of the nodes at the end of the edge but in reverse order. In this way, it is possible to visualize easily connections between different layers. See~\cite{BAHB2008} for further details of the visualization.}
\label{fig:0}
\end{figure}

Real heterogeneous networks are typically hierarchically organized. One of the most useful tools to uncover such hierarchies is the $k$-core decomposition~\cite{Dorogovtsev:2006ik}. Given a network, its $k$-core is defined as the maximal subgraph such that all nodes in the subgraph have at least $k$ connections with members of the subgraph. This defines a hierarchy of nested subgraphs, where the $1$-core contains the $2$-core, which in turn contains the $3$-core and so on until the maximum $k$-core is reached. Nodes belonging to the $k$-core but not to the $(k+1)$-core are said to have coreness $k$. Real networks often show a deep and complex $k$-core structure, as made evident by tools such as LaNet-vi~\cite{BAHB2008}. However, even though clustering has been shown to induce strong $k$-core hierarchies\cite{Serrano:2006dq}, the $k$-core {\it per se} does not include any information about clustering and, thus, cannot discriminate well between two networks with different global organization of clustering but with the same clustering coefficient.

To overcome this problem, the concept of $k$-core has been remodeled to account for clustered networks. A key ingredient throughout the paper is the concept of edge multiplicity $m$, defined as the number of distinct triangles going through an edge~\cite{Radicchi:2004av,Serrano:2006qj}. All edges belonging to a clique of size $n$ have identical multiplicity $n-2$ whereas an edge connecting two cliques has zero multiplicity.  
Therefore, strong correlations between the multiplicities of adjacent edges indicate that triangles are arranged in a clique-like fashion whereas a weaker correlation indicate a random distribution of triangles. It is therefore clear that, in order to uncover the global organization of triangles in a network, it is necessary to understand the organization of the multiplicities of their edges. This can be achieved with the $m$-core, defined as the maximal subgraph such that all its edges have, at least, multiplicity $m$ within it. This concept was developed in~\cite{Saito:2008nx,Gregori2013213} under the name of $k$-dense decomposition. The edges in a $k$-dense graph have multiplicity $m=k-2$. 
Because of this, we prefer the notion of $m$-core, which is directly related to the multiplicity: an edge belongs to the $m$-core if its multiplicity within the $m$-core is, at least, $m$. A node belongs to the $m$-core if at least one of its edges belongs to it. A node belonging to the $m$-core but not to the $(m+1)$-core is said to have $m$-coreness $m$. 
As in the case of the $k$-core, the $m$-core defines a set of nested subgraphs whose properties informs us about the global organization of triangles in the graph. The left plot in Fig.~\ref{fig:0} shows an example of a simple network and its $m$-core structure.

In the case of the $k$-core, the density of links within each subgraph grows as $k$ is increased. As a consequence, it is very unlikely that the $(k+1)$-core is fragmented in different components if the $k$-core is connected. Therefore, the main interest of the $k$-core decomposition is focused on the size of the giant $k$-core and the maximum coreness of the system. The situation is completely different in the case of the $m$-core. This is so because of a weaker correlation between $m$-coreness of a node and its degree~\cite{Orsini:2013}. In fact, the $m$-core decomposition is able to distinguish between a strong hierarchical structure --when $m$-cores do not fragment into smaller components-- from a highly modular architecture --when $m$-cores are always fragmented. In this case, the quantities of interest are, besides the size of the giant $m$-core and the maximum $m$-coreness, the number of components as a function of $m$.

\begin{figure}[t]
\centerline{\includegraphics[width=\linewidth]{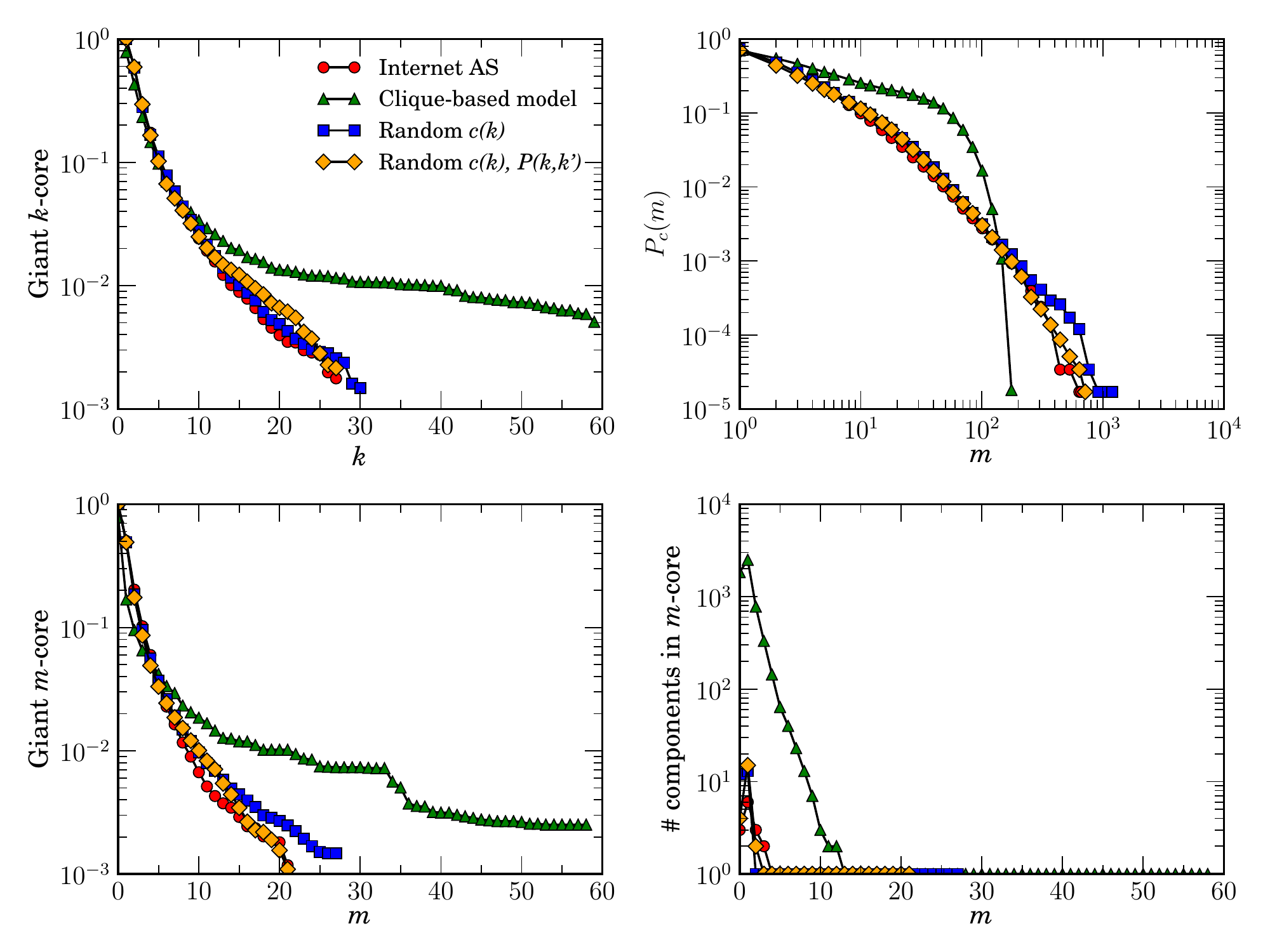}}
\caption{{\bf Measuring hierarchies in real and random networks.} Comparison of the $k$-core and $m$-core decompositions between the real Internet AS network, the clique based model, and maximally random models. ``Random $c(k)$'' stands for the maximally random model with a fixed degree distribution and clustering spectrum $c(k)$. ``Random $c(k)$, $P(k,k')$'' stands for the maximally random model that preserves also the degree-degree correlation structure of the real network. The top left plot shows the relative size of the giant $k$-core as a function of $k$. Top right plot shows the complementary cumulative distribution of edge multiplicities. Bottom left plot shows the relative size of the giant $m$-core as a function of $m$.  Finally, the bottom right plot shows the number of components in the $m$-core as a function of $m$.}
\label{fig:1}
\end{figure}

\begin{figure}[t]
\centerline{\includegraphics[width=\linewidth]{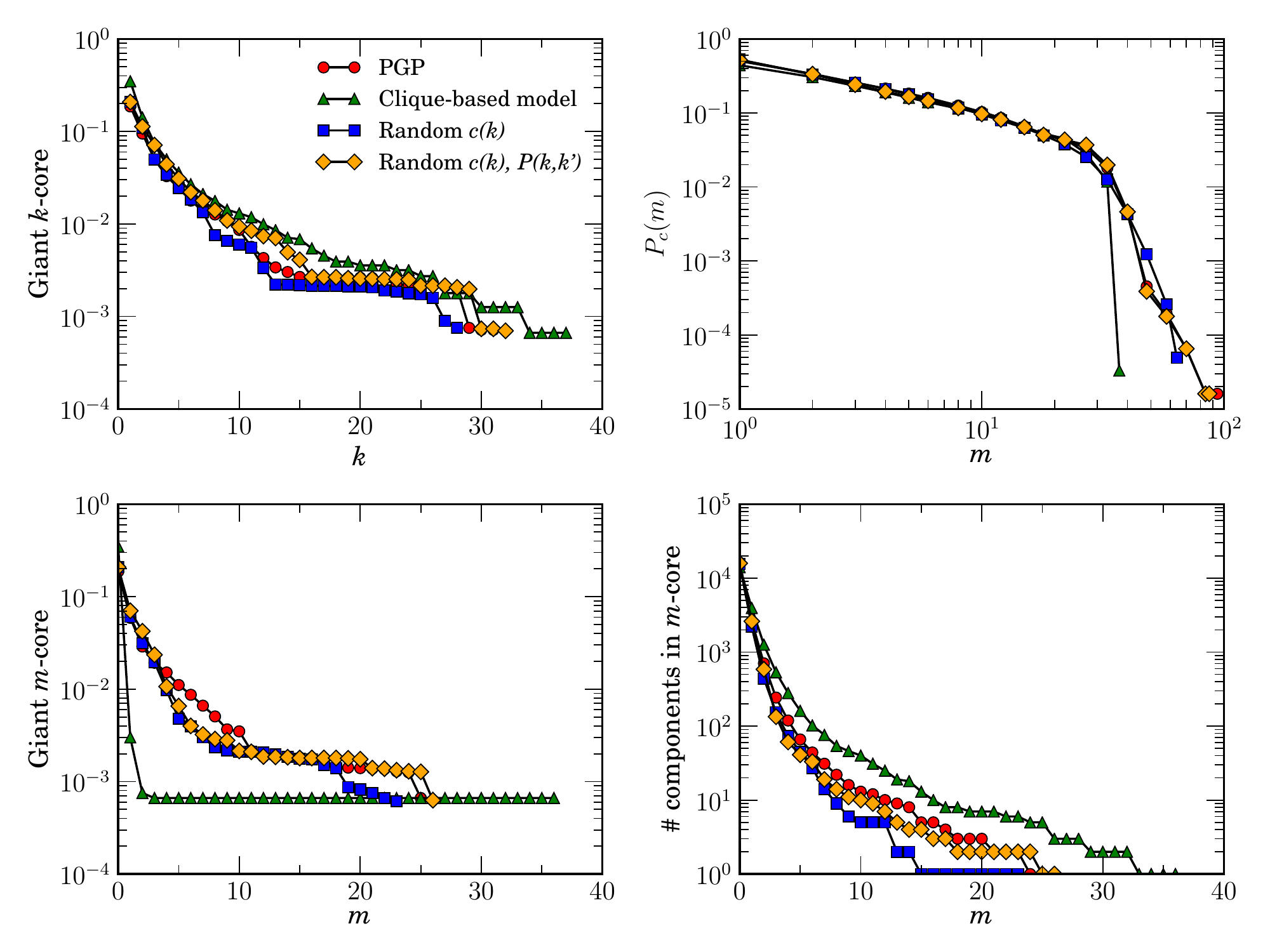}}
\caption{{\bf Measuring hierarchies in real and random networks.} The same as in Fig.~\ref{fig:1} but for the PGP web of trust.}
\label{fig:2}
\end{figure}

\begin{figure}[t]
\centerline{\includegraphics[width=\linewidth]{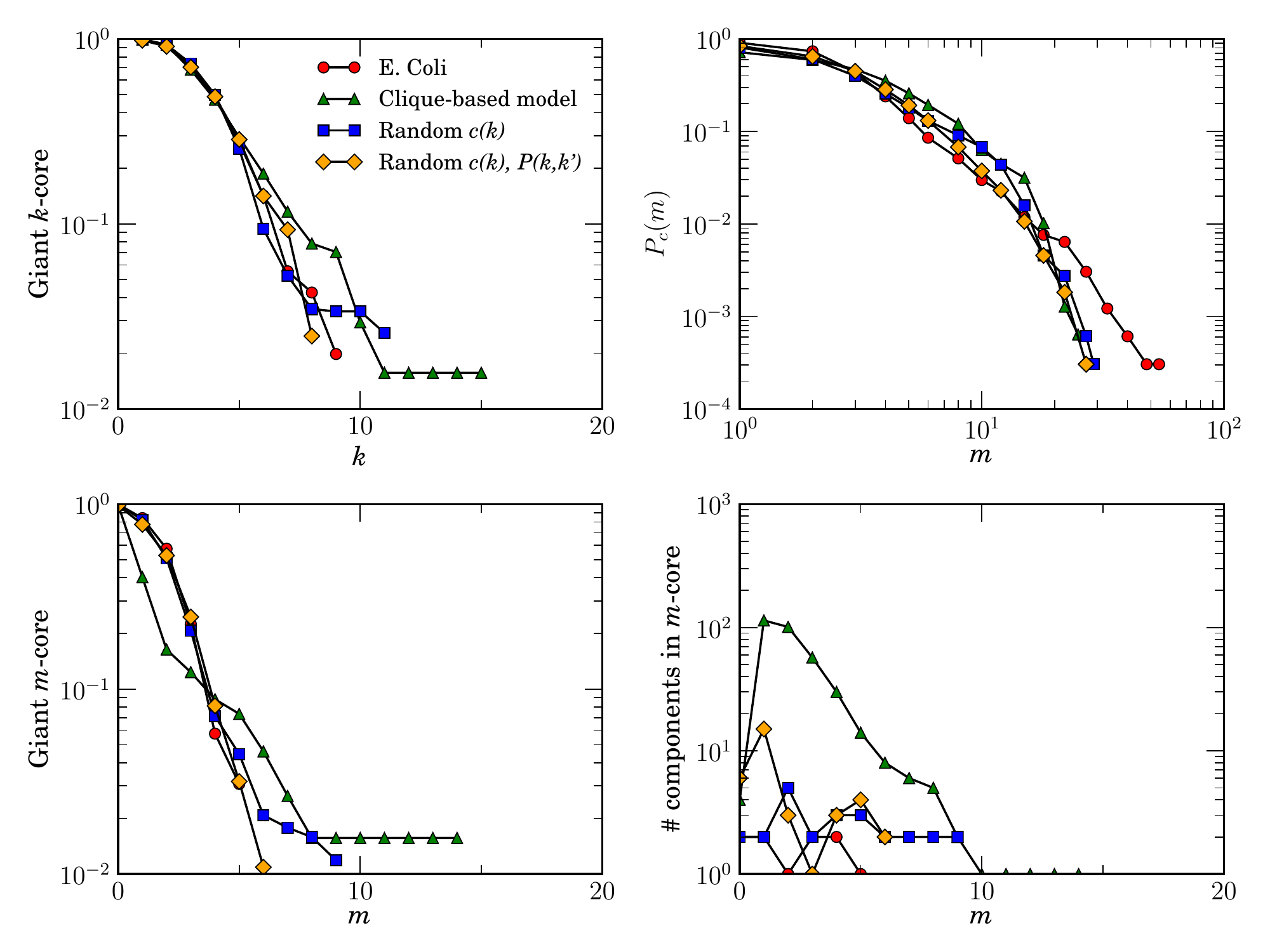}}
\caption{{\bf Measuring hierarchies in real and random networks.} The same as in Fig.~\ref{fig:1} but for the {\it E. Coli} metabolic network.}
\label{fig:3}
\end{figure}

Figures~\ref{fig:1}, \ref{fig:2}, and \ref{fig:3} show a comparison of the $k$-core and $m$-core decompositions between real networks and their random equivalents. As it can be observed in the top plots of these figures, all models do a reasonably good job at reproducing both the $k$-core structure and the distribution of edge multiplicities, even though MR models are clearly better than the CB one. However, there are important differences in the $m$-core decomposition. While both versions of MR models reproduce well the giant $m$-core, the maximum $m$-coreness, and the number of components as a function of $m$ of all the studied networks, the CB model overestimates the size and number of components in the case of the Internet and underestimate the size of giant $m$-cores in the PGP web of trust. In the case of the metabolic network, MR models reproduce well its entire $m$-core structure. The CB model, on the other hand, does not capture well the $m$-core decomposition. Even though the CB network is originally connected, it fragments into a large number of disconnected components already at the $m1$-core and keeps fragmenting at each level almost up to the largest $m$-core, which is also three times larger than the real one.

\begin{figure*}[t]
\centerline{\includegraphics[width=\linewidth]{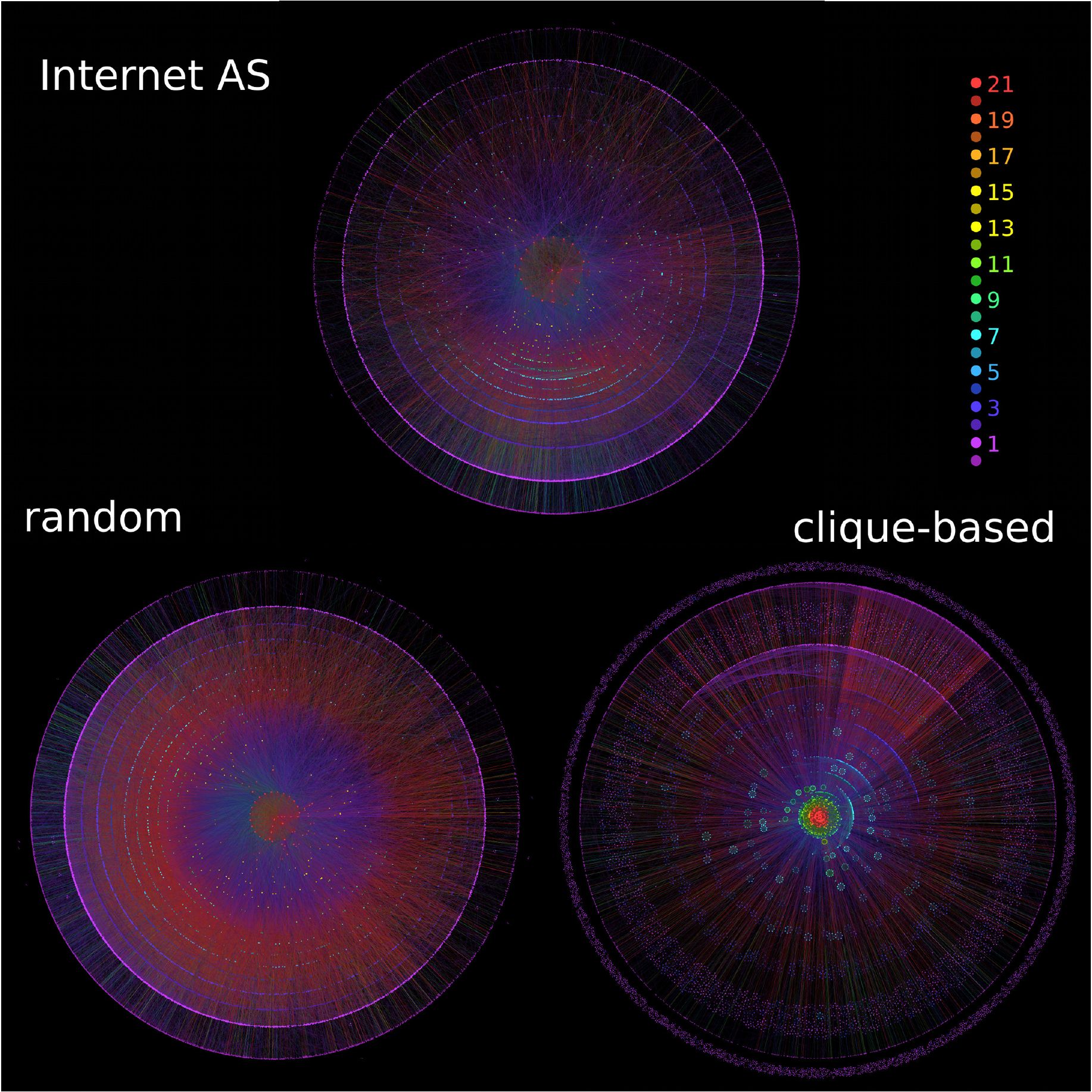}}
\caption{{\bf Visualizing $m$-cores.} $m$-core decomposition of the Internet AS network and its random versions. The MR version shown on the bottom left plot of the figure corresponds to the ``Random $c(k)$'' model, that is, with the rewiring scheme that does not preserves degree-degree correlations. The latter case is always closer to the real network. The color code is determined by the real network and kept the same in its random versions. However, layers in random networks above the maximum $m$-coreness of the real network are colored all in red. Maximum $m$-coreness for the MR and CB models are $27$ and $58$, respectively.}
\label{figInternet}
\end{figure*}

\begin{figure*}[t]
\centerline{\includegraphics[width=\linewidth]{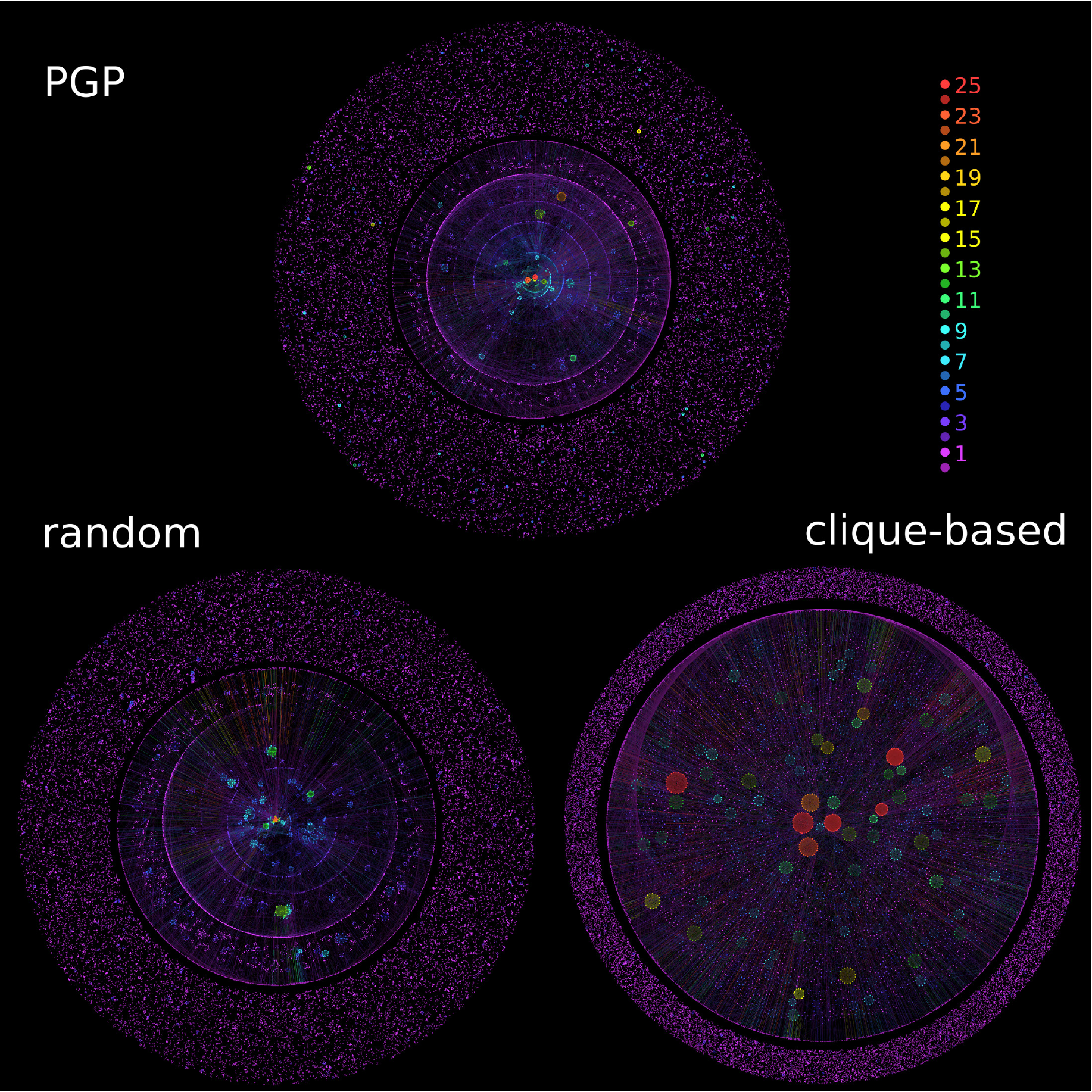}}
\caption{{\bf Visualizing $m$-cores.} The same as in Fig.~\ref{figInternet} for the PGP network and its random versions. Maximum $m$-coreness for the MR and CB models are $23$ and $36$, respectively.}
\label{figPGP}
\end{figure*}

\begin{figure*}[t]
\centerline{\includegraphics[width=\linewidth]{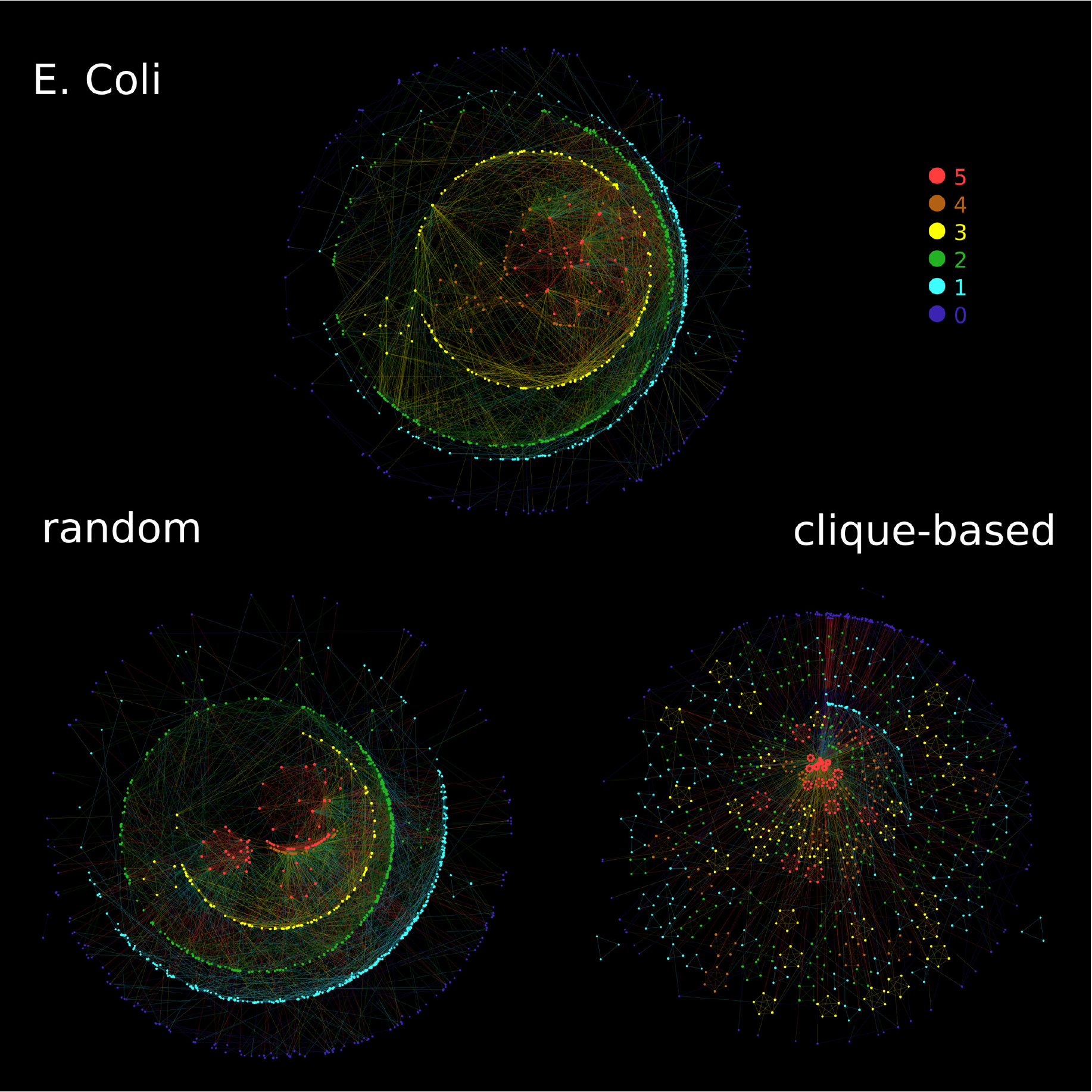}}
\caption{{\bf Visualizing $m$-cores.} The same as in Fig.~\ref{figInternet} for the {\it E. Coli} metabolic network and its random versions. Maximum $m$-coreness for the MR and CB models are $9$ and $14$, respectively.}
\label{figEcoli}
\end{figure*}

\subsection{$m$-core visualization}

The $m$-core decomposition is actually much richer and complex than what Figs.~\ref{fig:1}, \ref{fig:2}, and \ref{fig:3} show. Certainly, the $m$-core decomposition can be represented as a branching process that encodes the fragmentation of $m$-cores into disconnected components as $m$ is increased. The tree-like structure of this process informs us about the global organization --for instance hierarchical vs. modular-- of clustering in networks. To visualize this process we use LaNet-vi 3.0~\cite{LaNet-vi3}, a modified version of LaNet-vi~\cite{BAHB2008}, originally designed to visualize the $k$-core structure of a network. In short, LaNet-vi tool evaluates the coreness of all nodes of the network and arrange them in a plane following the hierarchy induced by the $k$-cores, so that nodes with high coreness are placed at the center of the figure whereas nodes with lower coreness are located around nodes with higher coreness in an onion-like shape. The major modification in LaNet-vi 3.0 with respect to the visualization mode in the previous version  concerns the representation of disconnected components. If the network forms a single connected component, nodes with $m$-coreness 0 are arranged in the outermost circle of the representation. Whenever the $m1$-core is fragmented into several components, these are arranged in separate and non-overlapping disks within the circle of $m$-coreness 0, with nodes of $m$-coreness 1 placed at the edge of their corresponding disk. The process is repeated for each disconnected component with the $m2$-core, $m3$-core, etc., until the maximum $m$-coreness present in the network is reached. The size of each disk is proportional to the logarithm of the number of nodes in the component. In this way, it is possible to visualize simultaneously all the information encoded in the $m$-cores so that different networks can be easily compared (see the right plot in Fig.~\ref{fig:0} for a simple example). When the original network is already fragmented (like in the PGP web of trust, for instance), we first proceed to arrange disconnected components in non overlapping disks within the outermost disk, that in this case does not have any node in its perimeter.

Figures~\ref{figInternet},~\ref{figPGP}, and~\ref{figEcoli} show the visualization of $m$-cores of real networks and their random equivalents (visualizations of MR models are shown only for $P(k)$ preserving rewiring). In the case of the Internet graph, the $m$-core visualization reveals a strongly hierarchical structure, where each layer is contained within the previous layer and where connections are mainly radial, with nodes with low $m$-coreness connected to nodes with higher $m$-coreness and very few connections between nodes in the same layer. Interestingly, this type of structure is also revealed in recent embeddings of the Internet graph into the hyperbolic plane~\cite{Boguna:2010uq}. This structure is very well reproduced by MR models, as it can be seen in the left bottom plot of Fig.~\ref{figInternet}, but not by the CB model, which generates a highly modular and non-hierarchical structure. The case of the web of trust of PGP is particularly interesting. Figure~\ref{figPGP} reveals a mixture of a modular structure, with a strong fragmentation for all values of $m$ --as one would expect for a social network--, and a hierarchical structure, revealed by the existence of a persistent giant $m$-core and a large number of layers. Again, this structure is very well reproduced by MR models whereas the CB model generates a very flat modular structure without any hierarchy. Finally, the metabolic network is also strongly hierarchical, although due to the small network size the number of layers is relatively small. MR models reproduce very well its structure whereas the CB model does not generate any hierarchy.

\section{Discussion}

The results presented in this paper indicate that, in agreement with previous studies~\cite{jamakovic:2009,Foster:2011fk}, the degree distribution $P(k)$ and clustering spectrum $\bar{c}(k)$ are the main contributors to the global organization of the majority of real networks, which are close to maximally random once these properties are fixed. 
This supports the idea that most real networks are the result of a self-organized process based on local optimization rules, in contrast to global optimization principles, that yield a hierarchical organization that cannot be reproduced by maximally ordered clustered models. Besides, the strong clustering observed in real networks, supports also the idea that such local principles are related to a similarity measure among nodes of the network that can be quantified by an underlying metric structure~\cite{Serrano:2008hb,Boguna:2009uz,Boguna:2010uq,KrPa10,Serrano:2012we,Papadopoulos:2012uq}. On the other hand, global optimization principles are necessarily present, for instance, in power grids, where they induce topologies that are very different from what one would expect at random. This is made evident by its $m$-core decomposition (see Supplementary Information). In this case, even thought the $m$-core structure is not very deep, it is very different from any of the random models, which generate highly unstructured $m$-cores. Therefore, the $m$-core decomposition along with its visualization tool can help us to find the true mechanisms at play in the formation and evolution of real networks.

\section{Methods}
\subsection{Maximally random clustered networks}

Maximally random clustered networks are generated by means of a biased rewiring procedure. We use two different rewiring schemes. In the first one, two different edges are chosen at random. Let these connect nodes A with B and C with D. Then, the two edges are swapped so that nodes A and D, on the one hand, and C and B, on the other, are now connected. We take care that no self-connections or multiple connection between the same pair of nodes are induced by this process. This rewiring scheme preserves the degree distribution of the original network but not degree-degree correlations. In the second rewiring scheme, we first chose an edge at random and look at the degree of one of its attached nodes, $k$. Then, a second link attached to a node of the same degree $k$ is chosen and the two links are swapped as before. Notice that this procedure preserves both the degree of each node and the actual nodes' degrees at the end of the two original edges. Therefore, the procedure preserves the full degree-degree correlation structure encoded in the joint distribution $P(k,k')$. Both procedures are ergodic and satisfy detailed balance.

Regardless of the rewiring scheme at use, the process is biased so that generated graphs belong to an exponential ensemble of graphs $\cal{G} = \mit \lbrace G \rbrace$, where each graph has a sampling probability $P(G)\propto e^{-\beta H(G)}$, where $\beta$ is the inverse of the temperature and $H(G)$ is a Hamiltonian that depends on the current network configuration. Here we consider ensembles where the Hamiltonian depends on the target clustering spectrum of the real Network $\bar{c}(k)$ as
\begin{equation}
H = \sum_{k=k_{min}}^{k_c} |\bar{c}^*(k)-\bar{c}(k)|,
\end{equation}
where $\bar{c}^*(k)$ is the current degree-dependent clustering coefficient. We then use a simulated annealing algorithm based on a standard Metropolis-Hastings procedure. Let $G'$ be the new graph obtained  after one rewiring event, as defined above. The candidate network $G'$ is accepted with probability
\begin{equation}
p = \min{(1,e^{\beta [H(G)-H(G')]})} = \min{(1,e^{-\beta \Delta H})},
\end{equation}
otherwise, we keep the graph $G$ unchanged. We first start by rewiring the real network $200E$ times at $\beta=0$, where $E$ is the total number of edges of the network. This step destroys the clustering coefficient of the original network. Then, we start an annealing procedure at $\beta_0=50$, increasing the parameter $\beta$ by a $10\%$ after $100E$ rewiring events have taken place. We keep increasing $\beta$ until the target clustering spectrum is reached within a predefined precision or no further improvement can be achieved. 

\subsection{Computing $m$-cores}

To compute $m$-cores efficiently, we develop a new approach, different from the one in~\cite{Saito:2008nx,Gregori2013213}. We first map the original graph $G$ into a hypergraph $G^*$, where edges in $G$ become vertices in $G^*$ and where each triangle in the original graph is mapped into an edge (a $3$-tuple) in $G^*$. Then, by noticing that the degree of a vertex $v^*$ in $G^*$ equals the number of triangles associated to the original edge in $G$, it is possible to obtain the $m$-core just by computing the $k$-core of the same level in $G^*$. The complete description can be found in the Supplementary Information.

\begin{acknowledgements}
This work was supported by MICINN projects No.\ FIS2010-21781-C02-02 and BFU2010-21847-C02-02; {\it Generalitat de Catalunya} grants No.\ 2009SGR838 and 2009SGR1055; the Ram\'on y Cajal program of the Spanish Ministry of Science; and by the ICREA Academia prize, funded by the {\it Generalitat de Catalunya}.
It was also supported by Argentine MINCyT project PICT-Bicentenario 01108, and UBACyT 2012 (20020110200181) of the {\it Universidad de Buenos Aires}.
\end{acknowledgements}

\end{document}